\documentclass[prb,aps,twocolumn]{revtex4}
\usepackage{graphicx}
\usepackage{epstopdf}
\usepackage{latexsym}
\usepackage{amsmath}
\usepackage{graphicx}
\usepackage{dcolumn}
\usepackage{bm}
\usepackage{wasysym}

\newcommand{\beq}{\begin{equation}}
\newcommand{\eeq}{\end{equation}}
\newcommand{\beqn}{\begin{eqnarray}}
\newcommand{\eeqn}{\end{eqnarray}}

\newcommand{\ua}{\uparrow}
\newcommand{\da}{\downarrow}

\begin{document}

\title{The Renormalization Group Studies on Four Fermion Interaction Instabilities on Algebraic Spin Liquids}

\author{Cenke Xu}
\affiliation{Department of Physics, Harvard University, Cambridge
MA 02138, USA}

\date{\today}

\begin{abstract}
We study the instabilities caused by four fermion interactions on
algebraic spin liquids. Renormalization group (RG) is used for
three types of previously proposed spin liquids on the square
lattice: the staggered flux state of SU(2) spin system, the
$\pi-$flux state of SU(4) spin system, and the $\pi-$flux state of
SU(2) spin system. The low energy field theories of the first two
types of spin liquids are QED3 with emerged SU(4) and SU(8) flavor
symmetries , the low energy theory of the $\pi-$flux SU(2) spin
liquid is the QCD3 with SU(2) gauge field and emergent Sp(4)
(SO(5)) flavor symmetry. Suitable large-$N$ generalization of
these spin liquids are discussed, and a systematic $1/N$ expansion
is applied to the RG calculations. The most relevant four fermion
perturbations are identified, and the possible phases driven by
relevant perturbations are discussed.
\end{abstract}

\maketitle

\section{Introduction}

Many algebraic spin liquid states have been proposed in 2+1
dimensional strongly correlated electronic systems. In these spin
liquids neither the spin rotation symmetry nor the spatial
discrete symmetry is broken, and the physical order parameters
have algebraic correlations. The gapless excitations of the system
include fractionalized spin excitations (the spinon) which are
usually centered around isolated Dirac points, and in many cases
also gapless gauge bosons. It is believed that when the number of
gapless spinons with Dirac fermion spectrum is small enough, or
when all the spinons are gapped, the gauge fields are confining.
However, with large enough fermion numbers $N$, the system is
believed to be described by a conformal field theory (CFT), with
the fixed point gauge field coupling $e^{\ast 2} \sim 1/N$.
Physics based on this conformal field theory at large-$N$ case has
been studied in many references
\cite{Hermele2005,hermele2004a,wen2002,ran2006}, and it has been
shown that the order parameters of various spin ordered patterns
with different symmetry breaking can be all described as fermion
bilinears at these critical spin liquids.

Reference \cite{wen2002a} has provided us with a general formalism
of studying the algebraic spin liquids. For spin-1/2 system, the
lattice mean field variational Hamiltonian enjoys a SU(2) local
gauge symmetry, on top of spin SU(2) global symmetry and all the
lattice symmetries. The specific type of gauge symmetry that
survives at low energy field theory depends on the choice of
background mean field variational parameters, and the low energy
gauge symmetries can be SU(2), U(1), or even $Z_2$. And since the
$Z_2$ gauge field only introduces short range interaction between
slave particles, the low energy long distance physics will not be
modified by $Z_2$ gauge field. Thus we will only consider spin
liquids with SU(2) and U(1) gauge field. Three types of spin
liquids are of particular interest to us. The first two are the
so-called staggered flux state of SU(2) spin system, and the
$\pi-$flux state of SU(4) spin system \cite{brad1988,brad1989}.
Both states are expected to be described by the following action
\beqn L = \sum_{a = 1}^{4N} \bar{\psi}_a\gamma_\mu(\partial_\mu +
ia_\mu)\psi_a + \cdots \label{action1}\eeqn The ellipses include
all the other terms allowed by the symmetry or generalized
symmetry transformations of the system. $\gamma_\mu$ with $\mu =
1$, 2, 3 are just three Pauli matrices, which is special for $d =
2$. Without the ellipses, action (\ref{action1}) describes a
conformal field theory. Spin liquids described by (\ref{action1})
enjoy U(1) local gauge symmetry and SU(4N) flavor symmetry. For
the staggered flux phase of SU(2) spin system, $N = 1$, and for
the $\pi-$flux phase of SU(4) spin system $N = 2$. Since the
SU(4N) flavor symmetry is larger than the physical symmetry, it is
called the emergent flavor symmetry. As has been studied
previously, all the fermion bilinears are forbidden by symmetry
and projective symmetry transformations (PSG) \cite{Hermele2005},
the only allowed local field theory terms which break the emergent
flavor symmetry down to physical symmetries are four fermion
interaction terms. Four fermion interactions violate the conformal
invariance of actions (\ref{action1}), therefore it plays the role
of instabilities of the CFT. In the limit of $N \rightarrow +
\infty$, the scaling dimension of any four fermion terms is 4,
which is obviously irrelevant. At finite $N$, whether these four
fermion terms are relevant or not can be studied by explicitly
calculating the $1/N$ corrections to the bare scaling dimension,
and this will be one of the goals of the current paper.

Since all the four fermion terms are scalar under all the physical
symmetry transformations, they should be mixed under RG flow. For
the U(1) spin liquids described by (\ref{action1}), we will
consider three types of four fermion terms. The first type of four
fermion terms will preserve the SU(4N) emergent flavor symmetry.
Due to the Pauli matrices nature of the Dirac matrices
$\gamma_\mu$, there are only two terms in this category, and they
will mix at the first order of $1/N$ expansion. We will show that
these four fermion terms are likely to be irrelevant for even very
small $N$, i.e. they will not create any instability. The second
type of four fermion terms will break the SU(4N) symmetry down to
Sp(4N) symmetry, this perturbation alone will be relevant at the
CFT for small enough $N$, and it is likely that it will drive the
system to another fixed point which describes a spin liquid with
Sp(4N) symmetry and gapless U(1) gauge bosons \cite{cenke2007}. At
$d = 2$ there is only one term of this kind. The third type of
four fermion terms break the flavor symmetry down to
$\mathrm{SU(2N)\times SU(2)}$, and we will show that there is also
a relevant linear combination. However with the presence of both
the second and third type of four fermion terms, the symmetry of
the system is only $\mathrm{Sp(2N)\times U(1)}$.

The physical meaning of these symmetry breaking can be understood
as following: In the $N = 1$ case, the physical symmetry is SU(2)
($\sim$ Sp(2)) spin symmetry plus all the lattice symmetries. It
is also quite popular to interpret the four fold degenerate
valence bond solid (VBS) states as an O(2) vector with $Z_4$
anisotropy, and the $Z_4$ symmetry breaking is possibly irrelevant
at the critical point between Neel and VBS phases
\cite{senthil2004,senthil2004a,sandvik2007a}. In the algebraic
spin liquid formalism in this work, the VBS states are interpreted
as fermion bilinears, and indeed transform as a planar vector
under the U(1) group generator $\mu^z$. The $Z_4$ anisotropy of
the O(2) vector should involve very high order of fermion
interactions which are negligible at the CFT. Thus at the end of
the chain of symmetry breaking, the symmetry is $\mathrm{Sp(2)
\times U(1)}$, which is identical to the symmetry with the
presence of both $\mathrm{SU(2N)\times SU(2)}$ four fermion terms
and Sp(4N) four fermion terms ($N = 1$). Thus driven by the
$\mathrm{SU(2N)\times SU(2)}$ terms, the Sp(4N) fixed point is
surrounded by phases with smaller symmetries, some of the phases
will break the Sp(2N) spin symmetry, and some other phases may
break the U(1) symmetry (the enlarged discrete symmetry).
Therefore the Sp(4N) fixed point is a critical point (or
multicritical point) between phases breaking completely different
symmetries.

The $N = 2$ case corresponds to the $\pi-$flux state of SU(4) spin
system on the square lattice, and recent numerical results suggest
that the $\pi-$flux state is a good candidate of the ground state
of SU(4) Heisenberg model on the square lattice \cite{assaad2005}.
The SU(4) spin and pseudospin symmetry have been discussed in
spin-orbit coupled systems \cite{khaliullin2000,zhang1998} as well
as spin-3/2 fermionic cold atom system
\cite{wu2003,wu2006b,wu2005a}. In spin-3/2 cold atom systems,
since the particle density is very diluted, only the $s-$wave
scattering should be considered. In this case, without fine-tuning
any parameter, the system automatically enjoys Sp(4) (SO(5))
symmetry. And by tuning the ratio between the spin-2 scattering
channel and spin-0 scattering channel, one can reach a critical
point with SU(4) spin symmetry. In the spin-3/2 cold atom system
at the vicinity of the SU(4) point, all the four fermion terms
discussed above should exist as a perturbation to the $\pi-$flux
state.

The third type of spin liquid we will discuss is the $\pi-$flux
state of SU(2) spin system. This state is invariant under SU(2)
local gauge transformation even at low energy field theory
\cite{wen2002a}: \beqn L = \sum_{l = 1}^3
\bar{\psi}\gamma_\mu(\partial_\mu - i a^l_\mu \sigma^l)\psi +
\cdots \eeqn $\sigma^l$ with $l = 1, 2$, $3$ are three Pauli
matrices of the SU(2) gauge group. The flavor symmetry of this
state has been shown to be Sp(4) \cite{ran2006}. However, the
SU(2) gauge field formalism makes the spin SU(2) symmetry
inapparent \cite{wen2002a}. In reference \cite{ran2006}, in order
to make the SU(2) gauge symmetry and the SU(2) spin symmetry both
apparent, the authors had to double the number of fermion
components, but now the fermion multiplet suffers from a
constraint: $\psi^\ast$ and $\psi$ are related through a unitary
transformation. In order to do calculations without constraint, in
this paper we will first introduce a Majorana fermion formalism
for this $\pi-$flux state. In this formalism there are eight
components of Majorana fermions with 2 Dirac species each, and the
system enjoys an SO(8) flavor symmetry in the absence of gauge
fluctuations. The SU(2) $\simeq$ SO(3) gauge group as well as
Sp(4) $\simeq$ SO(5) flavor symmetry group are both subgroups of
the SO(8) group. The Neel and VBS order parameters still form a
vector representation of the SO(5) flavor group. In the large-$N$
generalization, the gauge group is still SU(2), and the flavor
symmetry is Sp(2N), $N = 2^{n}$, $n = 1$, 2, $\cdots$. The
large-$N$ generalization is applicable to the $\pi-$flux state of
Sp(2N) spin models with $N = 2^{n-1}$. Our calculations show that
the SO(5) invariant four-fermion terms are not going to introduce
any instability to the $\pi-$flux state, while the SO(5) breaking
terms are relevant perturbations.

Our large-$N$ calculations have used some algebras and identities
of SU(N), Sp(2N) Lie Algebras. The detailed analysis of the group
theory and algebras will be summarized in the appendix. In section
II and III we will study U(1) and SU(2) spin liquids respectively.
In our calculations $1/N$ is the only small parameter used for
expansion, and we do not assume $\epsilon = d -1$ to be small. Our
loop integrals and field propagators are calculated in $d = 2$,
and a rigorous $\epsilon$ expansion should involve a general $d$
calculations. However, at general dimensions there are many more
four fermion terms than the $d = 2$ case, simply because at $d =
2$ the three gamma matrices are Pauli matrices, the Fierz identity
reduces the number of four fermion terms significantly, which is a
very convenient property we want to make full use of. A formal
rigorous general $d$ calculation is possible, we will leave it to
the future study.

\section{spin liquids with U(1) gauge field}

\subsection{SU(4N) four fermion terms}

The low energy field theory of the staggered flux state of SU(2)
spin system and the $\pi-$flux state of SU(4) spin system are
proposed to be described by CFT in equation (\ref{action1}). Four
fermion interactions are one type of instabilities. As has been
mentioned in the introductory section, we will focus on three
types of four fermion terms. The first type contains two terms:
\beqn L_1 = \frac{g_1}{4N\Lambda}(\bar{\psi}_{a}\psi_{a})^2, \
L_1^\prime =
\frac{g^\prime_1}{4N\Lambda}(\bar{\psi}_{a}\gamma_\mu\psi_{a})^2.
\label{4fermi1}\eeqn Hereafter the bracket denotes the trace in
the Dirac space. The number $N$ and cut-off $\Lambda$ at the
denominator is to guarantee both terms are at order of $N$ and the
coefficients are dimensionless constants. In equation
(\ref{4fermi1}), $a = 1, \cdots 4N$ is flavor indices, and we will
focus on the case with $N = 2^{n-1}$.

$L_1$ and $L_2$ are the only two four fermion terms which are both
SU(4N) and Lorentz invariant. Throughout the paper we will only
consider four fermion terms with Lorentz invariance, partly
because a large class of interesting quantum critical points are
$z = 1$ theories with emergent Lorentz invariance; the Lorentz
symmetry breaking effects in the kinetic terms of equation
(\ref{action1}) have been considered in reference
\cite{Hermele2005}, and they were showed to be irrelevant. Several
other SU(4N) invariant terms can be written down, for instance
$(\bar{\psi}_a\psi_b)(\bar{\psi}_b\psi_a)$,
$(\bar{\psi}_a\gamma_\mu\psi_b)(\bar{\psi}_b\gamma_\mu\psi_a)$,
$\sum_{i = 1}^{(4N)^2 - 1}(\bar{\psi}_a T^i_{ab}
\psi_b)(\bar{\psi}_c T^i_{cd}\psi_d)$ and $\sum_{i = 1}^{(4N)^2 -
1}(\bar{\psi}_a \gamma_\mu T^i_{ab} \psi_b)(\bar{\psi}_c\gamma_\mu
T^i_{cd}\psi_d)$. Here $T^i$ are fundamental representations of
SU(4N) algebra. However, using the Fierz identity of $\gamma_\mu$
matrices, and the identity (\ref{4n5}) in the appendix, all these
terms can be written as linear combinations of $L_1$ and $L_2$.

We will calculate the RG equation for the linear and quadratic
order of the four fermion couplings. The first order corrections
from $1/N$ expansion will be calculated for the linear term, and
for the quadratic terms only the leading order of unity is
calculated. Notice that when $g_1 = g_1^\prime = 0$ the system is
at the CFT fixed point, so the point with zero four-fermion
coupling is always a fixed point, despite the fact that gauge
field fluctuations will generate effective four-fermion
interactions \cite{herbut2005}, the effects of these generated
effective four-fermion interactions are included in diagram E and
F of Fig. \ref{fig2}. At the CFT fixed point, the scaling
dimensions of fermion bilinears have been calculated elsewhere
\cite{Hermele2005,hermele2007}. For instance, \beqn
\Delta(\bar{\psi} T_a\psi) = 2 - \frac{64}{3(4N)\pi^2}, \cr\cr
\Delta(\bar{\psi}\psi) = 2 + \frac{128}{3(4N)\pi^2}.\eeqn These
two fermion bilinears belong to different representations of the
SU(4N) algebra, therefore their scaling dimensions should in
principle differ from each other. Notice that the scaling
dimension of conserved current $\bar{\psi}\gamma_\mu\psi$ and
$\bar{\psi}\gamma_\mu T_a\psi$ do not gain any corrections from
the $1/N$ expansion at this CFT fixed point, simply because the
conservation law requires their scaling dimensions to be exactly
2.

\begin{figure}
\includegraphics[width=2.8in]{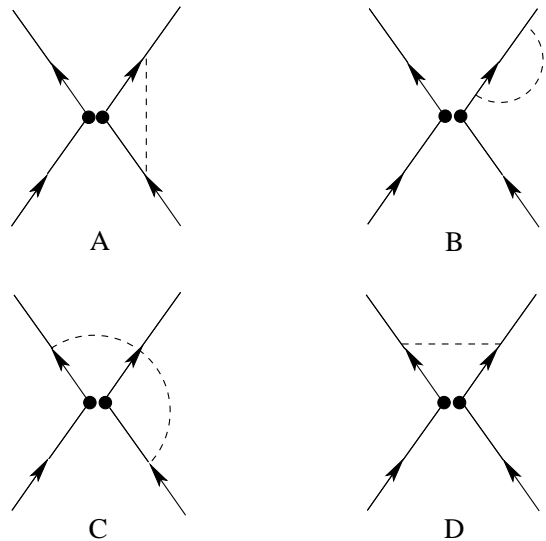}
\caption{Feynman diagrams contribute to the linear orders in both
Eqs. \ref{rgsun} and \ref{rgspn}. The dashed lines are dressed
photon propagators, and the full circles denote the trace in Dirac
space.} \label{fig1}
\end{figure}

\begin{figure}
\includegraphics[width=3.0in]{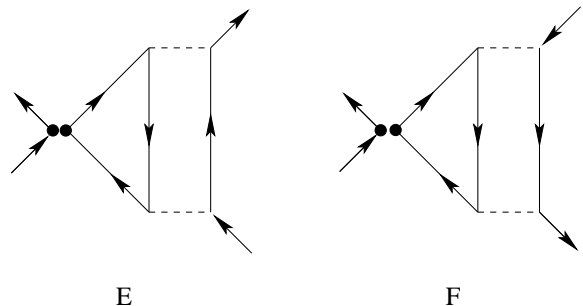}
\caption{Feynman diagrams which only contribute to the linear
orders in Eqs. \ref{rgsun}, but not in Eqs. \ref{rgspn}.}
\label{fig2}
\end{figure}

The $1/N$ correction of scaling dimensions mainly comes from the
dressed photon propagator \cite{lee1999}: \beqn G_{\mu\nu}(p) =
\frac{16}{4Np}(\delta_{\mu\nu} - \frac{p_\mu p_\nu}{p^2}).\eeqn
The Feynman diagrams which contribute to both $g_1$ and
$g_1^\prime$ are listed in Fig. \ref{fig1}. Diagrams A and B are
usually called the vertex and wavefunction renormalizations, which
also contribute to fermion bilinears. Besides these one-loop
diagrams, there are also two-loop diagrams (Fig. \ref{fig2}),
which involve two photon propagators and one extra trace in the
fermion flavor space, and hence also belongs to the $1/N$ order
correction.

\begin{figure}
\includegraphics[width=3.0in]{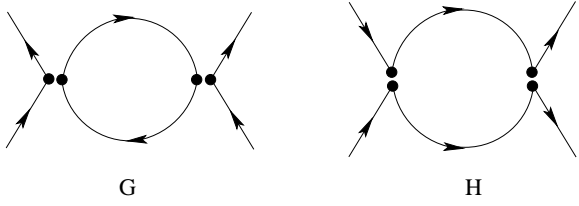}
\caption{Feynman diagrams which contribute to the quadratic order
of the RG equations (\ref{rgsun}) and (\ref{rgspn}). Notice that
since we only calculate to the order of unity in the quadratic
terms, diagram G only contributes to equation (\ref{rgsun}) but
not (\ref{rgspn}), and diagram H only contributes to equation
(\ref{rgspn}) but not (\ref{rgsun}).} \label{fig3}
\end{figure}

As already mentioned above, the quadratic terms in the equations
are only calculated to the order of unity, the only Feynman
diagram that contributes to this order is diagram G in Fig.
\ref{fig3}, all the other diagrams will contain one extra $1/N$.
After counting all the diagrams, the final RG equations are \beqn
\frac{dg_1}{d \ln l} &=& \left( - \epsilon - \frac{256}{3 (4N)
\pi^2} \right)g_1 + \frac{64}{4N \pi^2} g_1^\prime -
\frac{2}{\pi^2} g_1^2, \cr\cr\cr \frac{dg_1^\prime}{d \ln l} &=&
-\epsilon g^\prime_1 + \frac{64}{3 (4N) \pi^2} g_1 +
\frac{2}{3\pi^2} g_1^{\prime 2}. \label{rgsun}\eeqn Here $\epsilon
= d - 1 = 1$. At the fixed point $g_1 = g_1^\prime = 0$, the
largest eigenvalue of flowing equations is $- 1 + 1.39/(4N)$,
which is always negative for any integer $N$. Thus we conclude
that the The four fermion interactions which preserves SU(4N)
symmetries are likely irrelevant for all $N$. And no stable fixed
point is found at finite four fermion coupling. Another way of
interpreting this result is that, the flavor symmetry preserving
mass gap is not generated spontaneously, which is consistent with
reference \cite{herbut2005}.

\subsection{Sp(4N) four fermion terms}

The second type of four fermion terms will break SU(4N) symmetry
to Sp(4N) symmetry \cite{cenke2007}. In Sp(4N) algebra there is a
$4N\times 4N$ antisymmetric tensor $\mathcal{J}_{\alpha\beta}$
which satisfy \beqn \mathcal{J} T^a_{\mathrm{sp(4N)}} \mathcal{J}
= (T^{a}_{\mathrm{sp(4N)}})^t, \cr\cr \mathcal{J}
T^a_{\mathrm{su(4N)/sp(4N)}} \mathcal{J} = -
(T^{a}_{\mathrm{su(4N)/sp(4N)}})^t. \label{id1}\eeqn
$T^a_{\mathrm{sp(4N)}}$ are elements of Sp(4N) algebra, and
$T^a_{\mathrm{su(4N)/sp(4N)}}$ are elements in SU(4N) algebra but
not Sp(4N) algebra. All the algebra elements for $N = 2^{n-1}$
have been constructed in the appendix. The only four fermion term
of this type is \beqn L_2 =
\frac{g_2}{4N\Lambda}\mathcal{J}_{\alpha\gamma}\mathcal{J}_{\beta\sigma}
(\bar{\psi}_\alpha\psi_\beta)(\bar{\psi}_{\gamma}\psi_\sigma).\eeqn
The other current-current interaction term
$\mathcal{J}_{\alpha\gamma}\mathcal{J}_{\beta\sigma}
(\bar{\psi}_\alpha \gamma_\mu \psi_\beta)(\bar{\psi}_{\gamma}
\gamma_\mu \psi_\sigma)$ actually equals $L_2$ if one uses the
Fierz identity of Dirac gamma matrices in $d = 2$:
$\gamma^\mu_{\alpha\beta}\gamma^\mu_{\gamma\sigma} =
2\delta_{\alpha\sigma}\delta_{\beta\gamma} -
\delta_{\alpha\beta}\delta_{\gamma\sigma}$.

The Feynman diagrams in Fig. \ref{fig2} do not contribute to
$g_2$, and the diagram H in Fig. \ref{fig3} will contribute to the
order of unity in the quadratic term in the RG equation: \beqn
\frac{dg_2}{d\ln l} = \left( - \epsilon + \frac{64}{4N \pi^2}
\right)g_2 - \frac{1}{3 \pi^2} g_2^2. \label{rgspn} \eeqn This
equation has fixed points at $g_2=0$ and $g_2 = g_2^\ast = 3 \pi^2
(-\epsilon + 64/(4N \pi^2))$. At $d=2$ and $N=1$ we now find a
result which is very different from the SU(4N) perturbations
above. The $g_2=0$ fixed point is unstable with RG eigenvalue
$0.621$, while the fixed point at $g_2 = g_2^\ast >0$ is stable.
Notice that the quadratic term in this equation is the only term
with $\mathrm{O}(1/N^0)$ coefficient, all the other nonlinear
terms gain $1/N$ coefficient, thus the existence of this fixed
point can be obtained from $1/N$ expansion with $N$ extrapolating
back to $N = 1$, even without assuming $\epsilon$ to be small. All
the higher order terms in the $1/N$ expansions will only move the
critical point by order of $1/N$ at most. Although now the fixed
point value $g_2^\ast$ is of order unity, there is always a number
$4N$ at the denominator of $g_2$, thus $g_2/(4N)$ can still be
treated perturbatively close to the fixed point, as long as we do
not encounter an extra factor of $4N$ in our calculation. Because
$L_2$ is an pair-pair interaction term, no extra factor of $4N$ is
gained in our calculation if we only calculate the scaling
dimensions of terms like $\bar{\psi}T\psi$. The correction of
$g_2^\ast$ to the scaling dimensions of $L_1$ and $L_1^\prime$ is
also at the order of $1/(4N)$. For $N = 2$, to the order of
expansion done here, the fixed point with zero four-fermion terms
is stable against $L_2$ perturbation, and the finite four-fermion
coupling fixed point become instable. However, higher order $1/N$
corrections might change this result for $N = 2 $. Hereafter we
will denote the critical value of $N$ as $N_{c1}$. One can also
tune $N$ close to $N_{c1}$, and since $g_2^\ast$ is linear with
$(N - N_{c1})/N$, at the vicinity of the critical $N$, $g_2^\ast$
can be treated perturbatively.

\begin{figure}
\includegraphics[width=3.0in]{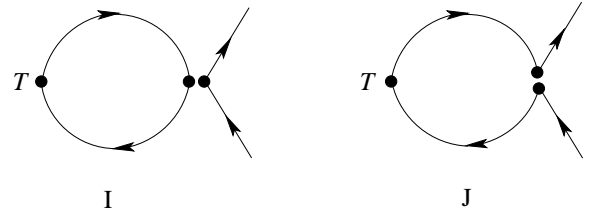}
\caption{Feynman diagrams which contribute to the difference of
scaling dimensions of fermion bilinears
$\bar{\psi}T^a_{\mathrm{su(4N)/sp(4N)}}\psi$ and
$\bar{\psi}T^a_{\mathrm{sp(4N)}}\psi$.} \label{fig4}
\end{figure}

The scaling dimensions of all $(4N)^2 - 1$ fermion bilinears
$\bar{\psi}T^a\psi$ ($T^a$ are SU(4N) generators) equal at the
fixed point with $g_i = 0$ which preserves the SU(4N) symmetry.
The difference between the scaling dimensions of
$\bar{\psi}T^a\psi$ and $\bar{\psi}\psi$ is from the diagrams
similar to the ones in Fig. \ref{fig3} \cite{hermele2007} with two
photon propagators and a trace in the fermion flavor space, which
only contributes to fermion bilinear $\bar{\psi}\psi$. At the
Sp(4N) symmetric fixed point, the scaling dimensions of fermion
bilinears are classified as the representation of Sp(4N) algebra:
$\bar{\psi}\psi$ and $\bar{\psi}T^a_{\mathrm{sp(4N)}}\psi$ form
scalar and adjoint representations of Sp(4N) group respectively,
the appendix proved that
$\bar{\psi}T^a_{\mathrm{su(4N)/sp(4N)}}\psi$ also form a
representation of Sp(4N) group at least for $N = 2^{n-1}$. For
instance, for the case with $n=1$, SU(4)/Sp(4) are just five Gamma
matrices, which form a vector representation of SO(5) (Sp(4))
algebra. The scaling dimensions of fermion bilinears within the
same representation are equal to each other.

If we assume $(N - N_{c1}) / N$ and $1/N$ are of the same order,
the scaling dimensions of the fermion bilinears at the Sp(4N)
fixed point deviate from their value at the SU(4N) fixed point at
the order of $1/N^2$, and requires a lot more calculations. But
their differences at $1/N^2$ order can be calculated readily from
diagrams in Fig. \ref{fig4}: \beqn
\Delta(\bar{\psi}T_{\mathrm{su(4N)/sp(4N)}}\psi) -
\Delta(\bar{\psi}T_{\mathrm{sp(4N)}}\psi) =
\frac{6g^\ast_2}{4N\pi^2}. \label{scaling}\eeqn To obtain these
results we have used the identities in equation (\ref{id1}).
Without assuming $N-N_{c1}$ to be small, the scaling dimensions of
fermion bilinears at the Sp(4N) fixed point can be calculated to
the $1/N$ order as: \beqn
\Delta(\bar{\psi}T_{\mathrm{su(4N)/sp(4N)}}\psi) = 2 -
\frac{64}{3(4N)\pi^2} + \frac{3g^{\ast}_2}{ 4N\pi^2}, \cr\cr
\Delta(\bar{\psi}T_{\mathrm{sp(4N)}}\psi) = 2 -
\frac{64}{3(4N)\pi^2} - \frac{3g^{\ast}_2}{ 4N\pi^2}, \cr\cr
\Delta(\bar{\psi}\psi) = 2 + \frac{128}{3(4N\pi^2)} +
\frac{3g^{\ast}_2}{ 4N\pi^2}. \label{sp4scaling}\eeqn Notice that
the diagrams in Fig. \ref{fig4} are the only two diagrams which
can contribute to the $1/N$ order of the scaling dimensions of
fermion bilinears.

The Sp(4N) fixed point is located at the side with $g_2 > 0$. At
the Sp(4N) fixed point, the modified linear order RG equation for
$L_1$ and $L_1^\prime$ reads: \beqn \frac{dg_1}{d \ln l} &=&
\left( - \epsilon - \frac{256}{3 (4N) \pi^2}  -
\frac{6g_2^\ast}{4N\pi^2} \right)g_1 + \frac{64}{4N \pi^2}
g_1^\prime, \cr\cr\cr \frac{dg_1^\prime}{d \ln l} &=& -\epsilon
g^\prime_1 + \frac{64}{3 (4N) \pi^2} g_1. \label{rgsun1}\eeqn Thus
at this Sp(4N) fixed point the SU(4N) perturbations $L_1$ and
$L_1^\prime$ are even more irrelevant compared with at the SU(4N)
fixed point.

When $g_2 < 0$ there is no stable fixed point and when $N$ is
small enough the system will be driven to a state with only short
range correlations. In this case the most favored state is likely
to be the Sp(4N) singlet pairing state. In general the pairing
amplitude $ \mathcal{J}_{\alpha\beta} \langle
\psi_{\alpha,i}\psi_{\beta,j} \rangle = C_{ij}$, and $C_{ij}$ is a
symmetric tensor, $i$, $j$ are Dirac matrices indices. For
convenience, we can choose the meanfield pairing amplitude to be $
\mathcal{J}_{\alpha\beta} \langle \psi_{\alpha,i}\psi_{\beta,j}
\rangle = C \delta_{ij}$, with constant $C$. This state breaks the
U(1) gauge symmetry to $Z_2$ gauge symmetry because of the fermion
pairing, the particle conservation of $\psi$ is also broken to
conservation mod 2, but the Sp(4N) symmetry is preserved simply
because the pairing is in the Sp(4N) singlet channel.

Using the identities (\ref{4n5}) proved in the appendix, $L_2$ can
also be written as \beqn L_2 = \sum_{a = 1}^{8N^2 - 2N -
1}\frac{g_2}{8N^2\Lambda}(\bar{\psi}T^a_{\mathrm{su(4N)/sp(4N)}}\psi)^2
+ \cdots \eeqn The ellipses are SU(4N) invariant terms $L_1$ and
$L_1^\prime$. Thus when $g_2$ is negative and grow large, the
system may also develop order
$\langle\bar{\psi}T^a_{\mathrm{su(4N)/sp(4N)}}\psi\rangle \neq 0$,
which breaks the Sp(4N) symmetry.
The competition between the Sp(4N) symmetry breaking state and the
Sp(4N) singlet pairing state requires further detailed analysis.

\subsection{$\mathrm{SU(2N)\times \mathrm{SU(2)}}$ four fermion terms}

The third type of four fermion terms are \beqn L_3 =
\frac{g_3}{4N\Lambda} (\bar{\psi}_{\alpha a}  \psi_{\alpha
b})(\bar{\psi}_{\beta b}  \psi_{\beta a}), \cr\cr L_3^\prime =
\frac{g^\prime_3}{4N\Lambda} (\bar{\psi}_{\alpha a} \gamma_\mu
\psi_{\alpha b})(\bar{\psi}_{\beta b} \gamma_\mu \psi_{\beta a}).
\eeqn Here $\alpha$ and $\beta$ are indices in the SU(2N)
subspace, and $a$ and $b$ are indices in the SU(2) space. These
two terms have other representations using identities proved in
the appendix: \beqn L_3 = \frac{g_3}{8N\Lambda}
(\bar{\psi}_{\alpha a}\vec{\mu}_{ab}\psi_{\alpha b})\cdot
(\bar{\psi}_{\beta c}\vec{\mu}_{cd}\psi_{\beta d}) + \cdots \cr\cr
L_3^\prime = \sum_{i = 1}^{(2N)^2 - 1} -
\frac{g_3^\prime}{4N^2\Lambda}(\bar{\psi}_{\alpha
a}T^i_{\alpha\beta}\psi_{\beta a})(\bar{\psi}_{\gamma
b}T^i_{\gamma\sigma}\psi_{\sigma b}) + \cdots \label{iden1}\eeqn
Again the ellipses are $L_1$ and $L_1^\prime$. The RG equations of
$g_3$ and $g_3^\prime $ will be mixed with $g_1$ and $g_1^\prime$
through the diagrams in Fig. \ref{fig2}. The final coupled RG
equations are \beqn \frac{dg_1}{d\ln l} = \left( - \epsilon -
\frac{256}{3 (4N) \pi^2} \right)g_1 + \frac{64}{4N
\pi^2}g_1^\prime \cr\cr\cr - \frac{64}{(4N) \pi^2}g_3 -
\frac{2}{\pi^2} g_1^2, \cr\cr\cr \frac{dg_1^\prime}{d \ln l} =
-\epsilon g^\prime_1 + \frac{64}{3 (4N) \pi^2} g_1 +
\frac{2}{3\pi^2} g_1^{\prime 2}, \cr\cr\cr \frac{dg_3}{d\ln l} =
\left( - \epsilon + \frac{128}{3 (4N) \pi^2} \right)g_3 +
\frac{64}{4N \pi^2} g_3^\prime - \frac{1}{\pi^2} g_3^2, \cr\cr\cr
\frac{dg_3^\prime}{d \ln l} = -\epsilon g^\prime_3 + \frac{64}{3
(4N) \pi^2} g_3 + \frac{1}{3\pi^2} g_3^{\prime 2}. \eeqn The
perturbation with the highest scaling dimension at the fixed point
with $g_i = 0$ is \beqn \lambda = -3/2 g_1 - 1/2 g_1^\prime + 3g_3
+ g_3^\prime, \label{lambda}\eeqn the scaling dimension is $-
\epsilon + 64/(4N \pi^2)$. When $N < N_{c2} = 64/(4\pi^2\epsilon)$
coupling constant $\lambda$ is clearly relevant, but when $N = 2$
at the first order calculation of $1/N$ expansion all the four
fermion terms are irrelevant, the highest scaling dimension is
about -0.189. However, higher order $1/N^2$ corrections might
change this result. The critical $N_{c2}$ we calculated is
consistent with the previous calculations in the context of
spontaneous chiral symmetry breaking mass generation of QED3
\cite{qed1984,qed1986,qed1988,herbut2005}.

Now let us assume $N < N_{c2}$, and after a long RG flow all the
irrelevant couplings are negligible. Thus at low energy and long
wavelength, $g_3 \simeq g_3^\prime$, $g_1 = g_1^\prime \simeq 0$,
and the relevant coupling $\lambda = 4g_3 = 4g_3^\prime$. Based on
equation (\ref{iden1}), positive relevant $\lambda$ tends to favor
SU(2N) symmetry breaking order $\langle \bar{\psi}
T^a_{\mathrm{su(2N)}} \psi \rangle \neq 0$, and negative relevant
$\lambda$ tends to favor SU(2) symmetry breaking order $\langle
\bar{\psi}\vec{\mu}\psi \rangle \neq 0$, which is usually referred
to as chiral symmetry breaking mass generation. In this case the
SU(4N) symmetric spin liquid becomes a critical point between two
phases with different symmetry breaking.

As is shown in the appendix, with the presence of all the four
fermion terms considered so far, the symmetry of the system is
broken down to $\mathrm{Sp(2N)\times U(1)}$, mainly because
neither the SU(2N) group nor the SU(2) group is a subgroup of
Sp(4N). In the case of $N = 1$ and staggered flux state, Sp(2N)
subgroup is the SU(2) spin symmetry, and U(1) is the effective
O(2) rotation of the planar vector formed by VBS order parameters.
In the case of $N = 2$ and $\pi-$flux state of SU(4) spin system,
realized in spin-3/2 cold atoms, Sp(2N) subgroup is the
unfine-tuned Sp(4) pseudospin symmetry.

\begin{figure}
\includegraphics[width=2.6in]{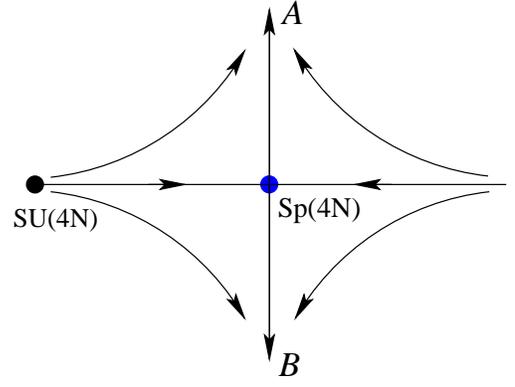}
\caption{The RG flow when $N < N_c = \mathrm{Min}[N_{c1},
N_{c2}]$, the horizontal axis is $g_2$, and the vertical axis is
$\lambda$ in (\ref{lambda}). At the Sp(4N) fixed point the order
parameters with equally strongest correlations are
$\bar{\psi}T^a_{\mathrm{sp(2N)}}\psi$ and $\bar{\psi}\mu^z\psi$,
thus phase $A$ is most likely to be order $\langle
\bar{\psi}T^a_{\mathrm{sp(2N)}}\psi \rangle $, and phase $B$ is
most likely to be $\langle\bar{\psi}\mu^z\psi\rangle$. Notice that
$\lambda$ is an RG eigenvector at the SU(4N) fixed point but not
at the Sp(4N) fixed point, but phase A and B can be driven by
$\lambda$ perturbation at the Sp(4N) fixed point.} \label{flow}
\end{figure}

At the first order $1/N$ expansion, two parameters have equally
the highest scaling dimensions: $g_2$ and $\lambda = - 3/2 g_1 -
1/2g_1^\prime + 3g_3 + g_3^\prime$, but the equality between the
two scaling dimensions are not protected by any symmetry. If $N <
N_c = \mathrm{Min}[N_{c1}, N_{c2}]$, both $g_2$ and $\lambda$ are
relevant, and $g_2$ is likely to drive the system to a fixed point
with Sp(4N) symmetry. Now let us focus on the vicinity of this
Sp(4N) fixed point. If we take $\epsilon$ of order unity, the
correction of the scaling dimension from fixed point value
$g_2^\ast$ will be at order $1/N$, and $L_3$, $L_3^\prime$ will be
mixing with many other terms with symmetry $ \mathrm{Sp(2N)
\otimes U(1)}$, the RG equations are rather complicated, but it is
very unlikely that there is no relevant flowing eigenvector.
Without detailed RG calculations, many results can be obtained
intuitively. Based on the identity (\ref{4n5}) proved in the
appendix we have: \beqn \sum_{a =
1}^{2N(4N+1)}(\bar{\psi}T^a_{\mathrm{sp(4N)}}\psi)
(\bar{\psi}T^a_{\mathrm{sp(4N)}}\psi) = \cr\cr
 - 2N \mathcal{J}_{\alpha\gamma}\mathcal{J}_{\beta\sigma} (\bar{\psi}_{\alpha}\psi_\beta)
 (\bar{\psi}_{\gamma}\psi_\sigma) + \cdots, \cr \cr \sum_{a = 1}^{8N^2 - 2N -1}
(\bar{\psi} T^a_{\mathrm{su(4N)/sp(4N)}}\psi)
(\bar{\psi}T^a_{\mathrm{su(4N)/sp(4N)}}\psi) = \cr\cr 2N
\mathcal{J}_{\alpha\gamma}\mathcal{J}_{\beta\sigma}
(\bar{\psi}_{\alpha}\psi_\beta)
 (\bar{\psi}_{\gamma}\psi_\sigma) + \cdots  \eeqn As was discussed in the previous paragraph, without $g_2$,
 relevant $\lambda$ tends to favor either SU(2N) symmetry breaking
 order $\langle\bar{\psi}T^a_{\mathrm{su(2N)}}\psi\rangle \neq 0$ or SU(2)
 symmetry breaking order $\langle\bar{\psi}\vec{\mu}\psi\rangle \neq
 0$, depending on the sign of $\lambda$. Notice that subalgebra $\mathrm{Sp(2N)\otimes 1}$
 and $1 \otimes \mu^z$ belong to Sp(4N), while
 $\mathrm{SU(2N)/Sp(2N)}$ and $1\otimes \mu^x$, $1\otimes \mu^y$
 all belong to SU(4N)/Sp(4N). A positive $g_2$ will favor order
 $\langle \bar{\psi}T^a_{\mathrm{sp(2N)}}\psi \rangle $
 over $\langle \bar{\psi}T^a_{\mathrm{su(2N)/sp(2N)}}\psi \rangle $ when $\lambda >
 0$; and also favors $\langle\bar{\psi}\mu^z\psi\rangle$ over
 $\langle\bar{\psi}\mu^x\psi\rangle$,
 $\langle\bar{\psi}\mu^y\psi\rangle$ with negative $\lambda$. Equation (\ref{sp4scaling}) also shows that
order parameter $\bar{\psi}T^a_{\mathrm{sp(2N)}}\psi $ and
$\bar{\psi}\mu^z \psi$ have stronger correlation and hence
stronger tendency to order at the Sp(4N) fixed point compared with
$\bar{\psi}T^a_{\mathrm{su(2N)}/\mathrm{sp(2N)}}\psi $ and
$\bar{\psi}\mu^x \psi, \ \bar{\psi}\mu^y \psi$. Therefore the
Sp(4N) fixed point is a critical point between Sp(2N) symmetry
breaking order
 $\langle \bar{\psi}T^a_{\mathrm{sp(2N)}}\psi \rangle$ and order
 $\langle\bar{\psi}\mu^z\psi\rangle$. The RG flow diagram is shown
 in Fig. \ref{flow}

From the first order $1/N$ expansion, $N_c$ is probably larger
than 1. In the case of staggered flux state with $N = 1$, the
theories above show that with the four fermion terms considered so
far, the Sp(4N) fixed point is a critical point between a SU(2)
symmetry breaking state, and a state which breaks time reversal
symmetry, but no spin or lattice translational symmetry is broken.
However, the SU(2) symmetry breaking state is not the Neel order.
The physical interpretations of these fermion bilinear order
parameters can be found in reference \cite{Hermele2005}. If the
critical number $N_c$ is also greater than 2, as in the case of
SU(4) $\pi-$flux state, the theory describes a critical point with
Sp(8) symmetry between an SO(5) symmetry breaking phase with order
parameter $\bar{\psi}\Gamma_{ab}\psi$, and a staggered chiral
state which breaks translational symmetry and time reversal
symmetry. Here $\Gamma_{ab} = \frac{1}{2i}[\Gamma_a, \Gamma_b]$
are spinor representations of SO(5) (Sp(4)) group, and $\Gamma_a$
($a$ = 1, 2 $\cdots$ 5) are five Gamma matrices.

When any of the fermion bilinear order is developed in the system,
the fermionic spectrum is gapped. In the case of gapped matter
field, the compact nature of the U(1) gauge boson is no longer
negligible, and the monopole proliferation usually opens a gap for
the gauge boson, and confine the gapped matter field. However, if
the gapped fermions form a topological insulator, the U(1) gauge
field is not necessarily confining. For instance, consider a Dirac
fermion system with conserved fermion charge coupled with a
compact U(1) gauge field, if the fermion gap $\bar{\psi}\psi$ is
turned on, the system enters the quantum Hall state of spinons,
and the Hall conductivity is half of the number of Dirac nodes. A
Chern-Simons term is generated for the compact U(1) gauge field,
in which case the monopole effect is suppressed
\cite{affleck1989}. This result can be understood physically as
following: a monopole in 2+1 dimensional space-time
annihilates/creates $2\pi$ flux of gauge field; however, an
adiabatically inserted $2\pi$ gauge flux will trap one spinon due
to the quantum Hall physics. And because of the conservation of
the spinon number, the $2\pi$ flux cannot be created or
annihilated freely.

In the order $\langle \bar{\psi}T^a_{\mathrm{sp(2N)}}\psi
\rangle$, the sign of the fermion gap, i.e. the sign of the Hall
conductivity depends on the Sp(2N) spin component. If a
$2\pi-$flux is adiabatically inserted in the system, it will trap
nonzero charge of $T^a_{\mathrm{sp(2N)}}$. In the past few years
the quantum spin Hall effect (QSHE) has attracted a lot of
attention, and many versions of QSHE models have been proposed,
most of which are two copies of quantum Hall states with opposite
Hall conductivities for spin up and down components
\cite{kane2005a,kane2005b}. Very recently the QSHE has been
observed in experiments \cite{zhang2006,zhang2007}. In our case
the state with $\langle \bar{\psi}T^a_{\mathrm{sp(2N)}}\psi
\rangle$ is actually a Sp(2N) generalization of a quantum spin
Hall model coupled with a compact U(1) gauge field. A nonzero
Sp(2N) spin will be trapped by an adiabatically inserted $2\pi$
gauge flux due to the QSHE effect. Because of spin conservation,
the monopole effect is again suppressed, thus in this state the
spinons are gapped but not confined. However, the stability of the
spin-filter edge states against the gapless U(1) gauge boson in
the bulk requires more careful analysis. This type of states will
be studied carefully in future \cite{xusenthil}. In the order
$\langle\bar{\psi}\mu^z\psi\rangle$, fermion gaps are opened for
two Dirac valleys with opposite signs, i.e. the total charge Hall
effect is zero. Also, since the O(2) rotation symmetry of $\mu^x$
and $\mu^y$ is broken down to $Z_4$ on the lattice, $\mu^z$ is not
precisely a conserved quantity. Therefore the flux tunnelling is
allowed, the monopoles are not suppressed, and the spinons are
still confined due to the compact nature of the U(1) gauge field.


We want to point out that we have not yet exhausted all the four
fermion terms allowed by the physical symmetry
$\mathrm{Sp(2N)\times U(1)}$. Terms like $(\bar{\psi}_{\alpha
a}\mu^z_{ab}\psi_{\alpha b})^2$,
$\mathcal{J}^\prime_{\alpha\gamma}\mathcal{J}^\prime_{\beta\sigma}
(\bar{\psi}_{\alpha a}\psi_{\beta a})(\bar{\psi}_{\gamma
b}\psi_{\sigma b})$ and some others are all allowed. Here
$\mathcal{J}^\prime$ is the antisymmetric tensor of the Sp(2N)
algebra, and in the appendix we will prove that $\mathcal{J} =
\mathcal{J}^\prime \otimes \mu^x$. Different four fermion terms
will favor different ordered patterns. For instance, if
$\mathrm{Sp(2N)\times U(1)}$ perturbation
$\sum_{b}(\bar{\mathrm{\psi}}T^b_{\mathrm{sp(2N)}}\otimes \mu^z
\psi)^2 - \sum_{a = x}^y (\bar{\psi}\mu^a\psi)^2$ flows to the
cut-off energy scale, it will drive a phase transition between the
Sp(2N) Neel and VBS order.

\section{spin liquid with SU(2) gauge field}

\subsection{The Majorana fermion formalism}

The $\pi-$flux state of SU(2) spin system on the square lattice
enjoys SU(2) local gauge symmetries. The low energy effective
theory of this state is \beqn L = \sum_{a = 1}^2\sum_{l = 1}^3
\bar{\psi}_a\gamma_\mu (\partial_\mu - ia^l_\mu\sigma^l)\psi_a +
\cdots \label{su2}\eeqn On the lattice, the variational parameter
$U_{ij}$ hopping matrix is chosen to be: $U_{i,i+\hat{x}} =
(-1)^{y} i\tau^0$, $U_{i,i+\hat{y}} = i\tau^0$, and the 2-site
unit cell is chosen to be $(i, i+\hat{y})$. For this choice of
gauge, the Dirac points are located at $(0, \pi/2)$ and $(\pi,
\pi/2)$. We use $a$ and $b$ to denote this two Dirac node valleys.
The Dirac gamma matrices are: $\gamma_0 = \sigma^2$, $\gamma_1 = -
\sigma^1$, $\gamma_2 = - \sigma^3$. It is believed that when the
fermion number is large enough, action (\ref{su2}) describes a
conformal field theory; when fermion number is small, the system
is instable against confinement due to the antiscreening
interaction between SU(2) gauge bosons
\cite{confine1990,confine1998}. In this section we will discuss
another type of instability of this conformal field theory driven
by four fermion interactions.

The SU(2) gauge field is operating on $\psi = (\psi_1, \psi_2)^T =
(f_{\ua}, - f_{\da}^\dagger)^T$. However, physical spin SU(2)
symmetry is not obvious in action (\ref{su2}). Since the charge
density in equation (\ref{su2}) is actually spin density $S^z$,
the charge current $\bar{\psi}\gamma_\mu\psi$ is not a singlet
under spin SU(2) transformation. In order to resolve this problem,
reference \cite{ran2006} enlarged the fermion space. However,
after this treatment there is a constraint on the fermionic space:
$\psi^\ast$ and $\psi$ are related through a unitary
transformation, this makes the calculations based on action
(\ref{su2}) inconvenient. In this section we will first introduce
a Majorana fermion formalism for the $\pi-$flux state. In this
Majorana fermion formalism both SU(2) gauge symmetry and SU(2)
spin symmetry are both apparent, and there is no constraint on the
fermion multiplet.

We define 8 component Majorana fermion multiplet $\chi$: \beqn
\chi_{111} = \mathrm{Re}(\psi_{1a}), \ \chi_{211} =
\mathrm{Im}(\psi_{1a}); \cr\cr \chi_{121} =
\mathrm{Re}(\psi_{2a}), \ \chi_{221} = \mathrm{Im}(\psi_{2a});
\cr\cr \chi_{112} = \mathrm{Re}(\psi_{1b}), \ \chi_{212} =
\mathrm{Im}(\psi_{1b}); \cr\cr \chi_{122} =
\mathrm{Re}(\psi_{2b}), \ \chi_{222} = \mathrm{Im}(\psi_{2b});
\eeqn Each index of $\chi$ denotes a two-component space. The
pauli matrices operating on the first, second and third two
component space are denoted by $\tau^a$, $\sigma^a$ and $\mu^a$
respectively. If we ignore gauge fluctuations, this system enjoys
an SO(8) symmetry, and all the symmetry transformations including
the SU(2) gauge transformations are subgroups of this SO(8) group.
We are going to write all the physical order parameters in terms
of bilinears of the Majorana fermion $\chi T \chi$, the fermion
statistics requires matrix $T$ to be antisymmetric.

Now we try to reformulate this theory in terms of $\chi$. In the
Majorana fermion space, the three $\mathrm{SU(2)}_{gauge}$
matrices are \beqn G_3 = \tau^2\otimes \sigma^x \otimes 1, \cr\cr
G_1 = \tau^2\otimes \sigma^z \otimes 1, \cr\cr G_2 = - 1 \otimes
\sigma^y \otimes 1. \label{gaugesu2}\eeqn One can check that these
three matrices, though mixing two different spaces, still form an
SU(2) algebra. All the physical symmetry transformation should
commute with this SU(2) algebra.

The bosonic version of our formalism actually realizes the
beautiful second Hopf map. One way to study the O(5) Nonlinear
Sigma model is to decompose the O(5) vector in terms of bosonic
SU(4) spinors as $n^a = \Phi^\dagger \Gamma_a \Phi$, $\Gamma_a$
with $a = 1$, 2, $\cdots 5$ are five Gamma matrices, and $\Phi$ is
a four component complex bosonic spinor \cite{demler1999}. After
this decomposition there is a redundant SU(2) gauge degrees of
freedom, and since the four component complex boson $\Phi$
contains eight real components, the effective field theory of O(5)
Nonlinear Sigma model with the Hopf term can be viewed as an O(8)
sigma model coupled with SU(2) gauge field. With unit length
constraint, the O(8) vector forms a manifold of seven dimensional
sphere $\mathrm{S^7}$, and the theory describes a mapping:
$\mathrm{S^7}/\mathrm{S^3} \rightarrow \mathrm{S^4}$, the
$\mathrm{S^3}$ manifold is exactly the SU(2) group manifold, and
$\mathrm{S^4}$ is the manifold formed by O(5) vector. This is a
direct generalization of the first Hopf map which gives the
$\mathrm{CP(1)}$ model, which is a popular way of rewriting the
O(3) Nonlinear Sigma model \cite{senthil2004}. This second Hopf
map has been used to construct the 4 dimensional quantum Hall
fluid \cite{zhang2001}. The Wess-Zumino-Witten term of the O(5)
Nonliear sigma model can also be derived from 2+1 dimensional
Dirac fermion action \cite{abanov2000}.

Inspired by the second Hopf map, the flavor symmetry of our theory
should be SO(5), in which the spin rotation symmetry should be
contained. After some algebras, one can see that the spin
transformation SU(2) algebra is \beqn S^z = - \tau^2\otimes 1
\otimes 1, \cr\cr S^x = \tau^1 \otimes \sigma^y \otimes 1, \cr\cr
S^y = \tau^3 \otimes \sigma^y \otimes 1. \label{spinsu2}\eeqn It
is straightforward to check that $[S^a, G_b] = 0$. The gauge group
generators in equation (\ref{gaugesu2}) and spin rotation
generators in equation (\ref{spinsu2}) together form an SO(4)
algebra.

There are in total 10 elements in the SO(8) algebra which commute
with the SU(2) gauge algebra, they are \beqn S^z = - \tau^2\otimes
1 \otimes 1, \ S^x = \tau^1 \otimes \sigma^y \otimes 1, \cr\cr S^y
= \tau^3 \otimes \sigma^y \otimes 1; \ - \tau^2 \otimes 1 \otimes
\mu^z, \cr\cr - \tau^2 \otimes 1 \otimes \mu^x; \ \tau^1 \otimes
\sigma^y \otimes \mu^z, \cr\cr \tau^1 \otimes \sigma^y\otimes
\mu^x, \ \tau^3 \otimes \sigma^y\otimes \mu^z, \cr\cr \tau^3
\otimes \sigma^y \otimes \mu^x, \ 1 \otimes 1\otimes \mu^y.
\label{su2so5}\eeqn These matrices are all antisymmetric, and form
an SO(5) algebra. Besides these antisymmetric matrices, there are
five symmetric matrices which form an vector representation of
this SO(5) algebra: \beqn \Gamma_1 = \tau^1 \otimes \sigma^y
\otimes \mu^y, \ \Gamma_2 = \tau^3 \otimes \sigma^y \otimes \mu^y,
\cr\cr \Gamma_3 = - \tau^2 \otimes 1 \otimes \mu^y, \ \Gamma_4 =
1\otimes 1 \otimes \mu^x,\cr\cr \Gamma_5 = 1 \otimes 1\otimes
\mu^z. \label{su2gamma}\eeqn The first three matrices form a
vector representation of spin SU(2) group, and it can be checked
that $[G_a, \Gamma_i] = 0$ for all $a$ and $i$. Now one can
construct fermion bilinears with SO(5) algebra constructed in
equation (\ref{su2so5}) and the Gamma matrices constructed in
equation (\ref{su2gamma}). The physical interpretation of all the
bilinears are summarized as following \beqn \mathrm{Neel}, \ n^a:
\bar{\chi} \Gamma_a \chi , \ a = 1, 2, 3; \cr\cr
\mathrm{ferromagnetic \ order}, \ m^a : \ \bar{\chi} \gamma_0 S^a
\chi; \cr\cr \mathrm{VBS_x}: \ \bar{\chi}\mu^x\chi, \
\mathrm{VBS_y}: \ \bar{\chi}\mu^z\chi; \cr\cr \mathrm{chirality}:
\ \bar{\chi}\chi; \cr\cr \mathrm{staggered} \ \mathrm{chirality}:
\ \bar{\chi}\gamma_0\mu^y\chi; \cr\cr (-1)^x \vec{S}_i \times
\vec{S}_{i+\hat{y}}: \ \bar{\chi}\gamma_0 S^a \mu^x \chi, \cr\cr
(-1)^y \vec{S}_i \times \vec{S}_{i+\hat{x}}: \ \bar{\chi}\gamma_0
S^a \mu^z \chi. \label{physical}\eeqn In the above equation,
$\bar{\chi} = \chi^T\gamma_0$. These bilinears have exhausted all
the elements in the SO(5) algebra and the $\Gamma_a$ matrices. All
these fermion bilinears correspond to long wavelength fluctuations
of certain order parameters on the lattice. The lattice version of
spin chirality is $S_1\cdot(S_2\times S_4) + S_2\cdot(S_3\times
S_1) + S_3\cdot(S_4\times S_2) + S_4\cdot(S_1\times S_3)$, $1$,
$2$, $3$ and $4$ are sites on the four corners of a unit square,
ordered clockwise.

The mean field choice of $U_{ij}$ apparently breaks the lattice
symmetry, thus the lattice symmetry transformations should be
combined with gauge transformations on the fermionic multiplet
$\psi$, which is usually called the projective symmetry group
(PSG). The complete PSG transformations combined with lattice
symmetry are summarized as: \beqn \mathrm{T_x}: \ 1\otimes
1\otimes \mu^z, \ \mathrm{T_y}: \ 1\otimes1\otimes \mu^x, \cr\cr
\mathrm{P_{xs}}: \ \gamma_1\otimes \mu^x, \ \mathrm{P_{xb}}: \
\gamma_1 \otimes i\mu^y, \cr\cr \mathrm{P_{ys}}: \ \gamma_2\otimes
\mu^z, \ \mathrm{P_{yb}}: \ \gamma_2 \otimes i\mu^y, \cr\cr
\mathrm{P_{xy}}: \ (\gamma_1 - \gamma_2)\otimes (\mu^x + \mu^z)/2,
\cr\cr \mathrm{T}: \ \gamma_0 \otimes i\sigma^y \otimes \mu^y.
\eeqn $\mathrm{T_x}$ and $\mathrm{T_y}$ are translations,
$\mathrm{P_{xs}}$ and $\mathrm{P_{ys}}$ are site centered
reflections, $\mathrm{P_{xb}}$ and $\mathrm{P_{yb}}$ are bond
centered reflections, and $\mathrm{P_{xy}}$ is reflection along
the line $x = y$, $\mathrm{T}$ is the time-reversal
transformation. The time reversal transformation is an antiunitary
operation, which transforms $i \rightarrow -i$. Therefore as long
as matrix $T$ between $\chi T \chi$ contains $i$, it always gains
an extra minus sign under time-reversal. For all the fermion
bilinears in equation (\ref{physical}), Neel order parameter,
ferromagnetic order parameter, chirality and staggered chirality
are odd under time-reversal; VBS order parameters and staggered
triplet bond order $(-1)^{i_\mu}\vec{S}_i\times
\vec{S}_{i+\hat{\mu}}$ are even.

It is interesting to compare the fermion bilinear representations
in the Majorana fermion formalism and the formalism in terms of
$\psi$. Introducing $\Psi = (\psi, -i\sigma^2\psi^\ast)^T$ and
$\bar{\Psi} = \Psi^\dagger\gamma^0$ as in reference
\cite{ran2006}, the comparison between fermion bilinears in the
$\chi$ language and $\psi$ language is listed below: \beqn 2
\bar{\chi}\chi = \bar{\Psi}\Psi, \ 2\bar{\chi} \Gamma_a \chi =
\bar{\Psi}\tilde{\Gamma}_a\Psi, \cr\cr 2\bar{\chi}\gamma_\mu
G_a\chi = \bar{\Psi}\gamma_\mu \sigma^a \Psi, \
2\bar{\chi}\gamma_\mu T^a \chi = \bar{\Psi}\gamma_\mu \tilde{T}^a
\Psi, \cr\cr \bar{\chi}\gamma_\mu\chi = \bar{\Psi}\gamma_\mu\Psi =
0, \ \bar{\chi}G_a\chi = \bar{\Psi}\sigma^a\Psi = 0, \cr\cr
\bar{\chi}T^a\chi = \bar{\Psi}\tilde{T}^a\Psi = 0. \eeqn $G_a$
with $a = 1, 2, 3$ are three matrices defined in equation
(\ref{gaugesu2}), and $T^a$ are ten SO(5) algebra generators
defined in equation (\ref{su2so5}). Notice that $\psi$ are
$-i\sigma^2\psi^\ast$ both transform as spinors under gauge SU(2)
group, $\sigma^a$ with $a = 1$, 2, 3 are three gauge SU(2) Pauli
matrices. The spin SU(2) transformation will mix $\psi$ and
$-i\sigma^2\psi^\ast$, the Dirac node valley space is another
direct product space. $\tilde{\Gamma}_a$ with $a = 1, 2, \cdots 5$
are five $4\times 4$ Gamma matrices operating on the spin space
and Dirac node valley space: \beqn \tilde{\Gamma}_1 =
\tilde{\sigma}^1 \otimes \mu^y, \ \tilde{\Gamma}_2 =
\tilde{\sigma}^2 \otimes \mu^y, \ \tilde{\Gamma}_3 =
\tilde{\sigma}^3 \otimes \mu^y, \cr\cr \tilde{\Gamma}_4 =
\tilde{1} \otimes \mu^x, \ \tilde{\Gamma}_5 = \tilde{1} \otimes
\mu^z, \eeqn and $\tilde{T}^a$ with $a = 1, 2, \cdots 10$ are
fundamental representations of ten $4\times 4$ $\mathrm{Sp(4)
\simeq SO(5)}$ generators, which are also the commutators of
$\tilde{\Gamma}_a$ matrices. Here $\tilde{\sigma}^a$ with $a = 1,
2, 3$ are three spin SU(2) Pauli matrices which mix $\psi$ and
$-i\sigma^2\psi^\ast$; $\mu^a $ with $a = x, y, z$ are three Pauli
matrices operating on the Dirac node valley space.

Now the field theory of $\pi-$flux state in terms of Majorana
fermions can be written as \beqn L = \sum_{l = 1}^3
\bar{\chi}\gamma_\mu(\partial_\mu - ia_\mu^l G_l)\chi + \cdots
\eeqn Here $\bar{\chi} = \chi^T\gamma_0$. The ellipses should
include all the four fermion terms allowed by PSG.

\subsection{large-$N$ generalization and RG equations for four-fermion perturbations}

The large-$N$ generalization of this problem can be achieved by
increasing two-component fermionic spaces. The gauge field always
only involves the first two 2-component spaces, and the gauge
group is always SU(2). The details of large-$N$ generalization is
in the appendix. Basically, for $n$ two-component fermionic
spaces, the number of Majorana fermions is $N_f = 2^n$, and the
flavor symmetry which commute with the SU(2) gauge algebra is
$\mathrm{Sp(4N)}$ with $N = 2^{n-3}$. All the matrices in the
particular representation of the Sp(4N) algebra are antisymmetric,
and there are $8N^2 - 2N - 1$ fermion bilinears
$\bar{\chi}\Gamma_a\chi$ which form a representation of Sp(4N)
algebra, $\Gamma_a$ are symmetric matrices. In the appendix we
also proved that our large-$N$ generalization corresponds to the
$\pi-$flux state of Sp(2N) spin system.

At the conformal field theory fixed point, the Majorana fermion
propagators are \beqn \langle \chi_{i,k}\bar{\chi}_{j,-k}\rangle =
\delta_{ij}\frac{ik_a\gamma_a}{2k^2}. \eeqn This can be viewed as
``half" fermion propagator. The dressed gauge field propagator
after integrating out the fermions is \beqn \langle
a^b_\mu(q)a^c_\nu(-q) \rangle =
\delta_{bc}\frac{32}{N_fq}(\delta_{\mu\nu} - \frac{q_\mu
q_\nu}{q^2}).\eeqn In our physical situation $N_f = 8$. The
scaling dimensions of some fermion bilinears can be calculated
readily: \beqn \Delta(\bar{\chi}\chi) = 2 + \frac{256}{\pi^2 N_f},
\cr\cr \Delta(\bar{\chi}\Gamma_a\chi) = 2 - \frac{128}{\pi^2 N_f},
\cr\cr \Delta(\bar{\chi}\gamma_\mu T^a\chi) = 2. \eeqn For all the
Majorana fermion bilinears, the matrix between $\chi$ should be
antisymmetric. One thing worth notice is that, the SU(2) gauge
current $\bar{\chi}\gamma_\mu G_a\chi$ is no longer gauge
invariant and hence has no well-defined scaling dimension. On the
contrary, the scaling dimension of SU(2) gauge singlet current
$\bar{\chi}\gamma_\mu T^a\chi$ gains no $1/N_f$ corrections.

The four fermion terms in the field theory should be invariant
under all the symmetry transformations, thus they should be mixed
under RG flow.
The simplest four fermion terms are squares of Sp(4N) scalar
fermion bilinears. To identify all the terms of this kind, we need
to find a symmetric tensor $\mathcal{T}$ or antisymmetric tensor
$\mathcal{J}$, which commute with gauge matrices $G^a$ and all the
Sp(4N) flavor matrices. If these tensors exist, one can write down
four fermion terms like \beqn (\bar{\chi}\mathcal{T}\chi)^2, \
(\bar{\chi}\gamma_\mu \mathcal{J}\chi)^2, \cr\cr \sum_{a = 1}^3
(\bar{\chi}\gamma_\mu \mathcal{T}G^a\chi)^2, \ \sum_{a = 1}^3
(\bar{\chi}\mathcal{J}G^a\chi)^2. \eeqn In the physical case with
$ N = 1$, the only symmetric tensor $\mathcal{T}$ one can find is
the unit matrix, and there is no satisfactory antisymmetric
$\mathcal{J}$. The representation of SO(5) in equation
(\ref{su2so5}) belongs to a vector representation of SO(8) group
and hence reducible, i.e. There are non-unit matrices commuting
with all the matrices in (\ref{su2so5}). However, the gauge
invariance criterion guarantees only the unit matrix $\mathcal{T}$
is satisfactory. Therefore the only two linear independent four
fermion terms of this type are \beqn L_1 =
\frac{g_1}{N_f\Lambda}(\bar{\chi}\chi)^2, \ L_1^\prime = \sum_{a =
1}^3 \frac{g_1^\prime}{N_f\Lambda}(\bar{\chi}\gamma_\mu
G_a\chi)^2. \label{su24fermi}\eeqn In the original $\psi$
language, these two terms are $(\bar{\psi}\psi)^2$ and $\sum_{\mu,
l}(\bar{\psi}\gamma_\mu\sigma^l\psi)^2$. Gauge singlet
Current-current interaction $(\bar{\chi}\gamma_\mu\chi)^2$ is not
allowed because of fermion statistics of $\chi$, i.e. in this
theory there is no extra global U(1) symmetry.
Both $L_1$ and $L_1^\prime$ are invariant under
$\mathrm{SU(2)_{gauge}\otimes Sp(2N)}$ group, and they are mixed
under RG flow at the linear order, i.e. the corrections from gauge
field fluctuations. The Feynman diagrams that contribute to the
anomalous dimensions are the same as the U(1) spin liquid case.
The coupled RG equations for $g_1$ and $g_1^\prime$ are \beqn
\frac{dg_1}{d\ln l} = ( - \epsilon - \frac{512}{\pi^2 N_f})g_1 +
\frac{384}{\pi^2 N_f}g_1^\prime - \frac{1}{\pi^2}g_1^2, \cr\cr\cr
\frac{dg_1^\prime}{d\ln l} = ( - \epsilon + \frac{1024}{3 \pi^2
N_f})g_1^\prime + \frac{128}{3\pi^2 N_f}g_1 +
\frac{1}{3\pi^2}g_1^{\prime 2}. \eeqn At the conformal field
theory fixed point, the most relevant combination is $\lambda_1 =
0.44 g_1 + g_1^\prime$, with scaling dimension $ - \epsilon +
36.5/N_f$, with critical $N_{f,c1} = 36.5$, which is much higher
than the spin liquids with U(1) gauge field fluctuation considered
in section II. At the physical case with $N_f = 8$, this conformal
field fixed point is very instable, and no stable fixed point is
found at finite four fermion couplings. The irrelevant RG flow
eigenvector is $-20.4 g_1 + g_1^\prime$, thus after long enough RG
flow, $g_1^\prime \simeq 20.4 g_1$, i.e. $L_1^\prime$ will
dominate $L_1$ at low energy and long wavelength, thus the phase
driven by these four fermion terms prefers to minimize
$L_1^\prime$. $L_1^\prime$ is a SU(2) gauge current interaction,
and gauge current $\bar{\chi}\gamma_\mu G_a\chi$ is not gauge
invariant. Therefore the order driven by $L_1^\prime$ can break
the SU(2) gauge symmetry. For instance, if the relevant flowing
eigenvector $\lambda_1$ is negative, it will flow to a state which
spontaneously generates a finite SU(2) gauge current on the
lattice scale, and this gauge current will break the SU(2) gauge
symmetry down to smaller gauge symmetries. If $\lambda_1 > 0$, the
possible state driven by $\lambda_1$ is a SU(2) gauge singlet
fermion paired state. Therefore if the Majorana fermion number $N$
is decreased from large enough value, two different instabilities
will compete: the SU(2) gauge boson confinement tends to drive the
system to an SU(2) gauge singlet ground state (the nature of this
phase is not clear); while the four fermion interaction studied in
the current work can drive the system to a state with broken SU(2)
gauge symmetry.

When any fermion bilinear order $\langle \bar{\chi}T\chi \rangle$
is developed, the fermion spectrum is gapped. The screening of
gapped fermions can no longer overcome the interactions between
SU(2) gauge bosons, the gauge coupling flow will confine all the
excitations with nonzero SU(2) gauge charge, all the excitations
of this phase have to be SU(2) gauge singlet. However, if the
SU(2) gauge symmetry is broken spontaneously by the relevant four
fermion terms, the residual gauge field fluctuation may or may not
be confining, depending on the gauge group. If the residual gauge
group is $Z_2$, the gapped spinons can be still deconfined; if the
residual gauge group is U(1), for instance when a uniform gauge
current $\bar{\chi}\gamma_\mu G^a \chi$ is generated, the monopole
fluctuation will confine the spinons, and the specific ground
state order pattern is determined by the quantum number of
monopoles. Notice that although both $L_1$ and $L_1^\prime$ are
SO(5) invariant, the SO(5) symmetry can be broken by the quantum
number of proliferating monopoles. A full analysis of the monopole
quantum number is not yet accomplished.

In the appendix we showed that the SU(2) gauge invariant formalism
is only exact for Sp(2N) Hamiltonian with $J_2 = 0$ in equation
(\ref{generalsp4}). When $J_2 \neq 0$ the system only enjoys the
U(1) gauge symmetry, and if $J_2 = J_1$ the spin model becomes
SU(2N) invariant, and the $\pi-$flux state is described by QED3.
Now let us consider turning on a small $J_2$ perturbation on the
$\pi-$flux state of Sp(2N) spin Hamiltonian with only $J_1$ in
equation (\ref{generalsp4}), this perturbation will generate
four-fermion perturbation \beqn L_2 = \frac{g_2}{N_f\Lambda} \{
2(\bar{\chi}\gamma_\mu G^3\chi)^2 - (\bar{\chi}\gamma_\mu
G^1\chi)^2 - (\bar{\chi}\gamma_\mu G^2\chi)^2  \}. \eeqn $L_2$ is
one component of the $d-$wave vector under gauge SU(2) group.
Since $L_2$ belongs to a different representation under SU(2)
gauge group from $L_1$ and $L_1^\prime$, the linear RG equation of
$L_2$ will not be mixed with $L_1$ and $L_1^\prime$. Also since
$\bar{\chi}G^a\chi$ vanishes due to fermion statistics, $L_2$
itself is an eigenvector under linearized RG flow to the first
order of $1/N$ expansion.

The RG equation for $L_2$ reads \beqn \frac{dg_2}{d \ln l} = (-
\epsilon + \frac{256}{3\pi^2 N_f})g_2 + \frac{1}{3\pi^2}g_2^2.
\eeqn Now the situation is similar to the $L_2$ considered in the
U(1) spin liquid case. The scaling dimension of $g_2$ is $-
\epsilon + 256/(3\pi^2 N_f)$, and for $N < N_{f,c2} = 256/(3\pi^2)
= 8.7$, $L_2$ will drive the system to a fixed point with a finite
$g_2$. At the fixed point of finite $g_2$, the SU(2) gauge
symmetry is broken down to U(1) gauge symmetry generated by $G^3$,
thus this fixed point is very analogous to the fixed point with
finite $L_2$ discussed in the section II. The critical value of
$N_{f,c2}$ from the first order $1/N_f$ expansion is slightly
larger than 8, and for the $\pi$-flux state of the Sp(4) spin
model with $N_f = 16$, $L_2$ will not bring an instability to the
state, and the finite $g_2$ fixed point becomes a critical point.

So far we have preserved the Sp(2N) flavor symmetry, which is
larger than the physical symmetry. As is discussed in the
appendix, our large-$N$ generalization is applicable to the
$\pi-$flux state of Sp(2N) spin model with $N = 2^{n - 1}$.
Therefore four fermion terms which break the emergent flavor
symmetry down to physical symmetries certainly exist in the field
theory. Let us assume the total number of 2-component fermion
space is $k+1$, two terms of this kind are \beqn L_3 =
\frac{g_3}{N_f \Lambda}\sum_{a} \{ 2(\bar{\chi}T^a_k \otimes \mu^y
\chi)^2 - (\bar{\chi}\mu^x \chi)^2 - (\bar{\chi}\mu^z \chi)^2]\},
\cr\cr L^\prime_3 = \frac{g^\prime_3}{N_f \Lambda}\sum_{a , b} \{
2(\bar{\chi}T^a_k \otimes \mu^y \gamma_\mu G^b \chi)^2 - \sum_{i =
x,z}(\bar{\chi}\mu^i\gamma_\mu G^b \chi)^2]\}. \cr \eeqn Notice
that fermion bilinear $\bar{\chi}T^a_{k }\otimes \mu^y \chi$ is
the large-$N$ analogue of the Neel order parameter,
$\bar{\chi}\mu^x \chi$ and $\bar{\chi}\mu^z \chi$ are the
large-$N$ analogue of the VBS order parameters, therefore a
relevant $L_3$ will favor either Neel or VBS phase depending on
the sign. The coupled linear RG equations for $g_3$ and
$g_3^\prime$ read \beqn \frac{dg_3}{d\ln l} = (-\epsilon +
\frac{256}{\pi^2 N_f})g_3 + \frac{384}{\pi^2 N_f}g_3^\prime,
\cr\cr \frac{dg^\prime_3}{d\ln l} = (-\epsilon + \frac{256}{\pi^2
N_f })g^\prime_3 + \frac{128}{3 \pi^2 N_f }g_3. \eeqn The most
relevant eigenvalue of the RG flow is $- \epsilon + 38.9/N_f$, the
critical value of $N_{f,c3}$ is $ 38.9$, which is slightly higher
than the critical value of $N_{f,c1}$ for $L_1$ and $L_1^\prime$
based on our first order $1/N$ expansion. If $N_{f,c3}$ is indeed
higher than $N_{f,c1}$, when $N_{f,c1} < N_f < N_{f,c3}$ the
$\pi-$flux state is a critical point between Neel order
$\bar{\chi}T^a_{k }\otimes \mu^y \chi$ and VBS order. The
classification of the four-fermion terms is worked on elsewhere
\cite{ranprivate}.

\section{concluding remarks}

In this work we studied the effects of four fermion interactions
as one type of instabilities on several interesting algebraic spin
liquids. The RG calculations show the gauge field fluctuation will
generally enhance the relevance of the four fermion interactions,
except for one particular pair which preserves the SU(4N) emergent
flavor symmetry in the spin liquids with U(1) gauge field. For the
$N = 1$ U(1) spin liquid, several four fermion terms are relevant
at the spin liquid. The four fermion term which breaks the SU(4N)
symmetry to Sp(4N) symmetry will likely drive the system to a
fixed point with finite coupling which describes a spin liquid
with Sp(4N) symmetry, which is a critical point between phases
with smaller symmetries. For the $N = 2$ case, all the four
fermion terms are irrelevant at the first order $1/N$ correction.
The $\pi-$flux state with SU(2) gauge field is more vulnerable
against four fermion terms, the critical fermion number is much
higher compared with U(1) spin liquids. The specific phases driven
by relevant four fermion couplings were conjectured in this paper,
but more detailed calculation is required to determine which
phases are most favorable ones.

Another physical system with low energy Dirac fermion excitations
is graphene, where the Dirac nodes locate at the corners of the
Brillouin zone. There are two flavors of Dirac fermions coming
from the two inequivalent corners of the Brillouin zone, and
another two flavors from the spin degeneracy. Thus in this system
the total number of Dirac fermions is $N = 4$. The difference
between this case and our spin liquids is that, there is no
fluctuating gauge field in graphene, except for a static Coulomb
interaction. Due to the apparent Lorentz symmetry breaking of the
Coulomb interaction, the fermi velocity will flow under RG. The
effects of four fermion terms in graphene have been studied in
reference \cite{son2007}.

It has been suggested that the deconfine critical point between
the Neel and VBS is of enlarged SO(5) symmetry
\cite{senthil2006,sandvik2007,kaul2007}, and the Neel and VBS
order parameters together form an SO(5) vector \cite{hu2005}. The
deconfined critical point between the Neel and VBS order is
conjectured to be a liquid phase of O(5) Nonlinear sigma model
with a Wess-Zumino-Witten term. A liquid phase with enlarged SO(5)
symmetry can exclude many possible relevant perturbations. In our
theory, SO(5) symmetry has appeared here and there, and both in
the U(1) spin liquids and the SU(2) spin liquid the Neel and VBS
order parameters form a five component SO(5) vector. Although we
have not completely identified the deconfine critical point in our
theory, our formalism especially the Majorana fermion formalism of
SU(2) $\pi-$flux state is still a promising approach to locate the
deconfine critical point, simply because of the beautiful second
Hopf map. To do this, one needs to find a fixed point with Sp(4)
flavor symmetry and only one relevant four fermion interaction
which breaks Sp(4) symmetry down to $\mathrm{SU(2)\otimes U(1)}$.
The fixed point with Sp(4) symmetry we identified in the U(1) spin
liquid section has one extra U(1) gauge symmetry compared with the
O(5) Nonlinear sigma model description of the deconfined critical
point \cite{senthil2006}, which in the dual language corresponds
to the conservation of gauge flux.

It is interesting to generalize the field theory of the deconfined
critical point to larger spin systems, and one can approach these
deconfined critical points from large-$N$ version of the spin
liquids studied in this work. First of all, the VBS order can be
naturally generalized to systems with Sp(2N) symmetry simply
because two Sp(2N) particles with fundamental representation can
form a Sp(2N) singlet through antisymmetric matrix $\mathcal{J}$:
$\mathcal{J}_{\alpha\beta}\psi^\dagger_{\alpha}\psi^\dagger_{\beta}$.
The Neel order parameter spans an adjoint representation of the
Sp(2N) group. The large-$N$ formalism of spin liquids in our
current paper shows that the smallest simple group with
$\mathrm{Sp(2N)\otimes U(1)}$ subgroup is Sp(4N). Therefore if an
unfine-tuned second order transition between Sp(2N) Neel and VBS
order exists, this critical point can enjoy enlarged Sp(4N)
symmetry.

\section{Appendix}

\subsection{Construction of fundamental representations of Sp(4N) algebra with $N = 2^n$}

In this appendix we will construct the fundamental representations
of SU(4N) and Sp(4N) algebras with $N = 2^n$. All the results will
be proved by induction, thus we will first present all the
results, which are obviously true for $n = 0$; later we will
assume they are also valid for $n = k$, the same results for $n =
k+1$ can be proved directly from our construction of SU(4N) and
Sp(4N) algebras.

{\bf 1st}, SU(4N) algebra contains subalgebra
$\mathrm{SU(2N)\otimes SU(2)}$, the whole fundamental
representation of SU(4N) algebra can be constructed from the
fundamental representations of its SU(2N) subalgebra and SU(2)
subalgebra. All the SU(4N) algebra elements can be written as
\beqn T_a \otimes \mu^i, \ T_a\otimes 1, \ 1\otimes \mu^i.
\label{su4n}\eeqn $T_a$ with $a = 1$, 2 $\cdots$ $(2N)^2 - 1$ are
fundamental representations of all the elements in SU(2N) algebra,
and $\mu^i$ with $i = 1$ 2, 3 are three SU(2) Pauli matrices.

{\bf 2nd}, SU(2N) algebra has an Sp(2N) subalgebra, which satisfy
\beqn \mathcal{J}_{\mathrm{2N}}
T^a_{\mathrm{sp(2N)}}\mathcal{J}_{\mathrm{2N}} =
(T^a_{\mathrm{sp(2N)}})^t. \eeqn Here $\mathcal{J}_{\mathrm{2N}}$
is a $2N \times 2N$ antisymmetric matrix.

{\bf 3rd}, all the SU(2N) elements in SU(2N)/Sp(2N) satisfy \beqn
\mathcal{J}_{\mathrm{2N}} T^a_{\mathrm{su(2N)/sp(2N)}}
\mathcal{J}_{\mathrm{2N}} = -(T^a_{\mathrm{su(2N)/sp(2N)}})^t.
\label{4n2}\eeqn

{\bf 4th}, all the elements in SU(2N)/Sp(2N) form a representation
of Sp(2N); or more precisely, fermion bilinear
$\bar{\psi}\Gamma^a_{\mathrm{su(2N)/sp(2N)}}\psi$ spans a
representation of Sp(2N) algebra. To prove this, one has to show
that \beqn [T^a_{\mathrm{sp(2N)}}, T^b_{\mathrm{su(2N)/sp(2N)}}]
\in \mathrm{SU(2N)/Sp(2N)}. \label{4n3}\eeqn The meaning of the
equation above is that, the commutator between any element in
SU(2N)/Sp(2N) and any element in Sp(2N) belongs to SU(2N)/Sp(2N).

{\bf 5th}, all the elements in SU(2N) algebra satisfy following
relations: \beqn [T^a_{\mathrm{sp(2N)}}, T^b_{\mathrm{sp(2N)}}]
\in \mathrm{Sp(2N)}, \cr\cr [T^a_{\mathrm{sp(2N)}},
T^b_{\mathrm{su(2N)/sp(2N)}}] \in \mathrm{SU(2N)/Sp(2N)}, \cr\cr
[T^a_{\mathrm{su(2N)/sp(2N)}}, T^b_{\mathrm{su(2N)/sp(2N)}}] \in
\mathrm{Sp(2N)}, \cr\cr \{ T^a_{\mathrm{sp(2N)}},
T^b_{\mathrm{sp(2N)}}\} \in \mathrm{SU(2N)/Sp(2N)}, \cr\cr \{
T^a_{\mathrm{sp(2N)}}, T^b_{\mathrm{su(2N)/sp(2N)}}\} \in
\mathrm{Sp(2N)}, \cr\cr \{ T^a_{\mathrm{su(2N)/sp(2N)}},
T^b_{\mathrm{su(2N)/sp(2N)}}\} \in \mathrm{SU(2N)/Sp(2N)}.\cr
\label{4n4}\eeqn

{\bf 6th}, for the fundamental representations of SU(2N) and
Sp(2N) algebras, following identities are satisfied: \beqn \sum_{a
= 1}^{(2N)^2 - 1}
T^a_{\mathrm{su(2N)},\alpha\beta}T^a_{\mathrm{su(2N)},\gamma\sigma}
= 2N\delta_{\alpha\sigma}\delta_{\beta\gamma} -
\delta_{\alpha\beta}\delta_{\gamma\sigma}, \cr\cr \sum_{a =
1}^{N(2N+1)}
T^a_{\mathrm{sp(2N)},\alpha\beta}T^a_{\mathrm{sp(2N)},\gamma\sigma}
\cr\cr = N\delta_{\alpha\sigma}\delta_{\beta\gamma} - N
\mathcal{J}_{\alpha\gamma}\mathcal{J}_{\beta\sigma}, \cr\cr
\sum_{a = 1}^{2N^2 - N - 1}
T^a_{\mathrm{su(2N)/sp(2N)},\alpha\beta}T^a_{\mathrm{su(2N)/sp(2N)},\gamma\sigma}
\cr\cr = N\delta_{\alpha\sigma}\delta_{\beta\gamma} + N
\mathcal{J}_{\alpha\gamma}\mathcal{J}_{\beta\sigma}-
\delta_{\alpha\beta}\delta_{\gamma\sigma}. \label{4n5}\eeqn All
these identities have been used in the main text of our paper.

Now the Sp(4N) algebra which is a subalgebra of SU(4N) can be
constructed as \beqn T^a_{\mathrm{sp(2N)}}\otimes \mu^x, \
T^a_{\mathrm{sp(2N)}}\otimes \mu^y, \ T^a_{\mathrm{sp(2N)}}\otimes
1, \cr\cr 1 \otimes \mu^z, \ T^a_{\mathrm{su(2N)/sp(2N)}}\otimes
\mu^z. \label{sp4n} \eeqn There are in total 2N(4N+1) elements in
(\ref{sp4n}). All these matrices satisfy \beqn
\mathcal{J}_{\mathrm{4N}} T^a_{\mathrm{sp(4N)}}
\mathcal{J}_{\mathrm{4N}} = (T^a_{\mathrm{sp(4N)}})^t , \cr\cr
\mathcal{J}_{4N} = \mathcal{J}_{\mathrm{2N}} \otimes \mu^x.
\label{4n1}\eeqn Meanwhile, All the elements in SU(4N) constructed
in (\ref{su4n}) but not in Sp(4N) constructed in equation
(\ref{sp4n}) are \beqn 1\otimes \mu^x, \ 1\otimes \mu^y, \cr\cr
T^a_{\mathrm{su(2N)/sp(2N)}} \otimes \mu^x, \
T^a_{\mathrm{su(2N)/sp(2N)}} \otimes \mu^y, \cr\cr
T^a_{\mathrm{su(2N)/sp(2N)}} \otimes 1, \ T^a_{\mathrm{sp(2N)}}
\otimes \mu^z \label{su4nsp4n} \eeqn There are in total
$\mathrm{8N^2 - 2N - 1}$ elements in equation (\ref{su4nsp4n}).

All the equations from (\ref{4n5}) to (\ref{4n1}) are valid for
SU(4) and Sp(4) algebras. Let us assume these results are true for
$n = k$, then for $n = k+1$, equations (\ref{4n1}), (\ref{4n2}),
(\ref{4n3}), (\ref{4n4}) and (\ref{4n5}) can be checked directly
through constructions in equations (\ref{su4n}), (\ref{sp4n}) and
(\ref{su4nsp4n}), and by using the assumptions made for $n = k$.
The calculations are tedious but straightforward.

We have proved that the fundamental representation of SU(4N)
algebra with $N = 2^n$ can all be constructed by Pauli matrices,
thus $( T^a_{\mathrm{su(4N)}} )^2 = 1$. Because of this and
equation (\ref{4n3}), vector $n^a =
(\bar{\psi}T^a_{\mathrm{su(4N)/sp(4N)}}\psi)$ rotates under Sp(4N)
group, while keeping the length $\sum_a (n^a)^2$ constant.

One can see that the SU(2N) subalgebra of SU(4N) does not
completely belong to Sp(4N) constructed in equation (\ref{sp4n}).
Instead, only the Sp(2N) subalgebra is a subalgebra of Sp(4N), and
the SU(2N)/Sp(2N) part belongs to SU(4N)/Sp(4N). Meanwhile, the
subalgebra SU(2) which commute with SU(2N) is not a subalgebra of
Sp(4N) either, only element $\mu^z$ which generates U(1) rotation
belongs to Sp(4N). Therefore when the $\mathrm{SU(2N)\otimes
SU(2)}$ and $\mathrm{Sp(4N)}$ four fermion terms both exist, the
symmetry of the system is actually only $\mathrm{Sp(2N)\otimes
U(1)}$.

\subsection{Large-$N$ generalization of the $
\pi-$flux state of SU(2) spin model}

In this subsection we will show that for Majorana fermions with
$n+1$ two-component space coupled with SU(2) gauge field, the
flavor symmetry is Sp(4N) with $N = 2^{n-2}$. In our paper we
showed that for $n = 1$ and $2$, the flavor symmetry is
SO(3)$\simeq$Sp(2) and SO(5)$\simeq$Sp(4) respectively, and for $n
= 2$ there are 5 symmetric matrices which make fermion bilinears
$\bar{\chi}\Gamma_a\chi$ span a representation of Sp(4). We will
try to generalize these results to larger number $n$. For $n = k$,
let us first denote the Sp(2N) algebra elements as $T^a_k$, and
denote the space spanned by the symmetric matrices as
$\Gamma_{\mathrm{sp(2N),k}}$; and second, assume for $n = k$,
following algebra is valid: \beqn [T^a_{k}, T^b_{k}] \in
\mathrm{Sp(2N)}_{k}, \ [\Gamma^a_{k}, \Gamma^b_{k} ] \in
\mathrm{Sp(2N)}_k, \cr\cr [T^a_{k}, \Gamma^b_{k}] \in
\Gamma_{\mathrm{sp(2N)},k}, \ \{ T^a_{k}, T^b_{k} \} \in
\Gamma_{\mathrm{sp(2N)},k}, \cr\cr \{ \Gamma^a_{k}, \Gamma^b_{k}
\} \in \Gamma_{\mathrm{sp(2N)},k}, \ \{ T^a_{k} , \Gamma^b_{k} \}
\in \mathrm{Sp(2N)}_{k}. \label{algebra}\eeqn These algebras are
valid for the simplest case with $n = 1$ and $n = 2$.

Now we construct Sp(2N) and $\Gamma_{\mathrm{sp(2N)}}$ for $n =
k+1$ as following: \beqn \mathrm{Sp(2N)}_{k+1}: \cr\cr T^a_{k}
\otimes \mu^x, \ T^a_{k}\otimes \mu^z, \ \Gamma^a_{k}\otimes
\mu^y, \ T^a_{k} \otimes 1, \ 1 \otimes \mu^y; \cr\cr
\Gamma_{\mathrm{sp(2N)},k+1}: \cr\cr \Gamma^a_{k}\otimes \mu^x, \
\Gamma^a_{k}\otimes \mu^z, \ T^a_{k}\otimes \mu^y, \ \Gamma^a_{k}
\otimes 1, \cr\cr 1 \otimes \mu^x, \ 1\otimes \mu^z.
\label{sok+1}\eeqn $\mu^a$ are Pauli matrices in the new two
component space. Although for this representation of Sp(2N)
algebra there is no antisymmetric matrices $\mathcal{J}$ which
satisfies $\mathcal{J}T^a \mathcal{J} = (T^a)^t$, the construction
in equation (\ref{sok+1}) is exactly the same as the construction
in (\ref{sp4n}) in the previous section, except for exchanging
$\mu^z$ and $\mu^y$, thus the algebra in equation (\ref{sok+1}) is
Sp(2N). There are in total $N(2N+1)$ elements in Sp(2N), and $2N^2
- N - 1$ elements in $\Gamma_{\mathrm{sp(2N)}}$. The validity of
algebra in (\ref{algebra}) for $n = k+1$ can be checked directly
by using the assumption (\ref{algebra}) and construction
(\ref{sok+1}). Notice that all the Sp(2N) elements in this
representation are antisymmetric and belong to a vector
representation of a larger group $\mathrm{SO}(2^{k+1})$, and the
fermion bilinear vector $\bar{\chi}\Gamma_a\chi$ rotates under
Sp(2N) group, with invariant vector length. The Sp(2N) algebra
constructed this way is the largest flavor symmetry commuting with
the SU(2) gauge algebra in equation (\ref{gaugesu2}).

In our calculation we have generalized the SU(2) $\pi-$flux state
to the case with larger number of fermion flavors by increasing
the number of 2-component fermion space. What kind of lattice
model can the large-$N$ generalization be applied to? Recall that
the smallest spin group SU(2)$\simeq$ Sp(2), one way of
generalizing SU(2) spin system is to generalize the spin symmetry
to Sp(2N), and let us assume $N = 2^n$. The lattice spin
Hamiltonian reads \beqn H = \sum_{<i,j>} J S^a_{i}S^a_j,
\label{latticespn}\eeqn $S^a$ are $N(2N+1)$ Sp(2N) Lie Algebra
elements. Introducing spinon $f_\alpha$ in the usual way $S^a =
f^\dagger_{\alpha}T^a_{\alpha\beta}f_\beta$ with half-filling
constraint $f^\dagger_{i,\alpha}f_{i,\alpha} = N$, we can use the
fundamental representation constructed in equation (\ref{sp4n}) in
the appendix to rewrite the Hamiltonian (\ref{latticespn}) as
\beqn H = \sum_{<i,j>} N J
(f^\dagger_{i,\alpha}f_{i,\beta}f^\dagger_{j,\beta}f_{j,\alpha} -
\mathcal{J}_{\alpha\gamma}\mathcal{J}_{\beta\sigma}
f^\dagger_{i,\alpha}f_{i,\beta}f^\dagger_{j,\gamma}f_{j,\sigma}).
\eeqn The meanfield variational parameters are defined as \beqn
\chi_{ij} = \langle f^\dagger_{i,\beta}f_{j,\beta} \rangle , \
\eta_{ij} = \mathcal{J}_{\alpha\beta} \langle
f_{i,\alpha}f_{j,\beta} \rangle. \eeqn In the above Hamiltonian we
have performed suitable transformation to make $\mathcal{J} =
i\sigma^y \otimes 1_1 \otimes \cdots \otimes 1_n$, with $N = 2^n$.
After particle-hole transformation, we define fermion multiplet
$\psi_{1,\alpha} = (f_1, \cdots f_N)^T$, $\psi_{2, \alpha} = (
f^\dagger_{N+1}, \cdots f^\dagger_{2N})^T$. the meanfield
Hamiltonian can be written as \beqn H = \sum_{<i,j>} NJ \{
\psi^\dagger_{i, a, \alpha}U_{ij,ab}\psi_{j, b,\alpha} + H.c. +
\frac{1}{2}\mathrm{Tr}[U^\dagger_{ij}U_{ij}] \}, \cr\cr U_{ij} = i
\mathrm{Re}(\chi) + \mathrm{Im}(\chi)\tau^3 +
\mathrm{Re}(\eta)\tau^1 + \mathrm{Im}(\eta)\tau^2.
\label{largenaction}\eeqn The Hamiltonian (\ref{largenaction})
enjoys a same SU(2) local gauge symmetry as the SU(2) spin
meanfield Hamiltonian \cite{wen2002a}. The Sp(2N) generalization
of the $\pi-$flux state can also be found in reference
\cite{ran2006}, where an opposite logic was taken, the Sp(2N) spin
operators were constructed from fermionic spinons.

The meanfield choice of variational parameters is the same as the
SU(2) $\pi-$flux state: $U_{i,i+\hat{x}} = (-1)^{y} i\tau^0$,
$U_{i,i+\hat{y}} = i\tau^0$, and the 2-site unit cell is chosen to
be $(i, i+\hat{y})$, the rest of the formulation is the same as
the SU(2) spin case, and the SU(2) gauge symmetry is preserved in
the low energy field theory. The flavor symmetry of the low energy
field theory action of the Sp(2N) $\pi-$flux state without
four-fermion terms should include the U(1) rotation between the
two Dirac nodes. Using the results in appendix $A$, it is
straightforward to show that the smallest simple group with $
\mathrm{Sp(2N)\otimes U(1)}$ subgroup is Sp(4N) with $N = 2^n$.
Therefore our large-$N$ generalization is applicable to the
$\pi-$flux state of Sp(2N) spin system with $N = 2^n$.

The SU(2) gauge symmetry of (\ref{largenaction}) is only exact for
Sp(2N) spin Hamiltonian (\ref{latticespn}). However, equation
(\ref{latticespn}) is not the only way to write down a nearest
neighbor Sp(2N) Hamiltonian. The general Hamiltonian reads \beqn H
= \sum_{<i,j>} J_1 T^a_iT^a_j + J_2\Gamma^a_i\Gamma^a_j, \cr\cr
T^a \in \mathrm{Sp(2N)}, \ \Gamma^a \in \mathrm{SU(2N)/Sp(2N)}.
\label{generalsp4}\eeqn The lattice SU(2) gauge symmetry is only
exact when $J_2 = 0$. When $J_1 = J_2$ the system enjoys the
SU(2N) spin symmetry, and it is known that the $\pi-$flux state of
SU(2N) system only has U(1) gauge symmetry when $N > 1$. Thus if
we turn on a small $J_2$ perturbation at the $\pi-$flux state, it
will induce four-fermion terms breaking the SU(2) gauge symmetry.

\begin{acknowledgments}
The author thanks S. Sachdev, Xiao-Gang Wen and Ying Ran for
helpful discussions. This work is supported by the Milton Funds of
Harvard University.

\end{acknowledgments}


\bibliography{4fermi}
\end{document}